\documentclass[prl,aps,tightenlines,floatfix,showpacs,twocolumn]{revtex4}
\usepackage{graphicx}
\usepackage{bm}
\usepackage{amsmath}
\usepackage{amssymb}
\begin{document}
  \title{Role of the Pauli principle in collective-model coupled-channels 
    calculations}
  \author{L. Canton$^{(1)}$}
  \email{luciano.canton@pd.infn.it}
  \author{G. Pisent$^{(1)}$}
  \email{gualtiero.pisent@pd.infn.it}
  \author{J. P. Svenne$^{(2)}$}
  \email{svenne@physics.umanitoba.ca}
  \author{D. van der Knijff$^{(3)}$}
  \email{dirk@unimelb.edu.au}
  \author{K. Amos$^{(4)}$}
  \email{amos@physics.unimelb.edu.au}
  \author{S. Karataglidis$^{(4)}$}
  \email{kara@physics.unimelb.edu.au}
  \affiliation{$^{(1)}$ Istituto  Nazionale  di  Fisica  Nucleare,  
    Sezione  di Padova, \\  e Dipartimento di Fisica  dell'Universit\`a 
    di Padova, via Marzolo 8, Padova I-35131, Italia}
  \affiliation{$^{(2)}$ Department  of  Physics  and Astronomy,  
    University  of
    Manitoba,   and  Winnipeg   Institute  for   Theoretical  Physics,
    Winnipeg, Manitoba, Canada R3T 2N2}
  \affiliation{$^{(3)}$ Advanced  Research   Computing,  Information  
    Division, University of Melbourne, Victoria 3010, Australia}
  \affiliation{$^{(4)}$ School  of Physics,  University of  Melbourne, 
    Victoria 3010, Australia}
  \date{\today}
  \begin{abstract}
 A multi-channel algebraic scattering theory, to find solutions of
 coupled-channel scattering problems with interactions determined by
 collective models, has been structured to ensure that the Pauli 
 principle is not violated. By tracking the results in the zero 
 coupling limit, a correct interpretation of the sub-threshold and 
 resonant spectra of the compound system can be made. As an example, 
 the neutron-${}^{12}$C system is studied defining properties of 
 ${}^{13}$C to 10~MeV excitation.  Accounting 
 for the Pauli principle in collective coupled-channels models is 
 crucial to the outcome. 
  \end{abstract}
  \maketitle


  At energies above 25 MeV, by using optical potentials formed by full
  folding effective two-nucleon interactions with microscopic (nucleon
  based)  descriptions  of the  target  structure,  the importance  of
  treating the Pauli principle  has been well established \cite{Am00}.
  However, in the domain of low-energy nucleon scattering for which an
  explicit  coupled-channels theory  of scattering  is  essential, the
  significance of  Pauli exclusion effects has not  been well defined.
  Many  coupled-channels codes  are available,  some of  which perform
  phenomenological   collective-model    calculations   searching   on
  parameter values of the chosen function  forms to find a best fit to
  experimental data.  But while it has  long been known  that any such
  models violate  the Pauli principle~\cite{Ma69,Gr96}, quantification 
  of that violation is lacking. 

  To  study  the effects  of  the  Pauli  principle in  a  macroscopic
  (collective model) approach is not  a trivial task.  In a recent 
  publication~\cite{Am03}, the orthogonalizing pseudo-potential (OPP) 
  method~\cite{Ku78,Sa69}  was generalized  to treat this 
  problem.  That was a small though important part of the full theoretical 
  framework of the multi-channel algebraic scattering (MCAS) theory of
  scattering~\cite{Am03}. Therein the OPP was used in finding the spectra, 
  bound and resonance properties, of ${}^{13}$C.  However, implications
  of the role of the Pauli principle in collective model coupled-channel 
  calculations arising from the use of the OPP was not discussed. Such is 
  a purpose of this letter.  Another is that the method could be pertinent 
  for any study requiring coupled channel solutions of quantal systems 
  involving fermions. As the example, we study the effects introduced by 
  the Pauli   principle  in   collective,  geometrical-type,   models  for
  low-energy  nucleon-nucleus processes  that  can  be characterized
  from  the spectrum of  the compound  nucleus. That spectrum includes  
  the states that lie below the  nucleon-nucleus threshold and in 
  the continuum as revealed by the narrow  and broad resonances that 
  lie  upon a smooth but
  energy dependent background of the elastic scattering cross section.
  This can be done in a systematic  and self-consistent way since the
  MCAS  approach  facilitates  such a  determination of  the sub-threshold 
  bound states and resonances of the compound nucleus.  This theory,  with  
  which  one   solves  the
  coupled-channel   Lippmann-Schwinger    (LS)   equations   for   the
  nucleon-nucleus system considered, is built upon sturmian expansions
  of an interaction matrix of potential functions.

  The  MCAS  method has  been  developed  in  momentum space  and  the
  starting  matrix of potentials  may be  formed by  folding effective
  two-nucleon  interactions  with  one-body  density matrices  of  the
  target  (studies  in  progress)  or,  as  is  more  common,  from  a
  collective model  description of the target  states and excitations.
  As in that recent publication \cite{Am03}, we have used a rotational
  collective-model  representation with  deformation  taken to  second
  order. We chose Woods-Saxon  functions and their various derivatives
  to be the  form factors for all components  each with characteristic
  operators of  diverse type. The interactions were  allowed to depend
  on parity  as well.  With such a  characterization, we were  able to
  describe all  important aspects, at positive  and negative energies,
  in the neutron-$^{12}$C system.

  With the MCAS  approach and a collective model  prescription for the
  starting matrix of potentials, the OPP is used in the process by which
  the  sturmians are specified. The OPP inclusion ensures that all 
  sturmians in the (finite) set selected as
  the basis of expansion of the matrix of potentials contain few or no
  components  equivalent to the  external nucleon  being placed  in an
  already densely  occupied orbit. That scheme is  an approximation as
  we discuss  later by  assessing the spectra  of $^{12,13}$C  and the
  single  neutron  spectroscopic   amplitudes  that  link  them  using
  information   obtained  from   large  space   no-core   shell  model
  calculations. But it is a good approximation.

  The  role of  the Pauli  principle is  studied by  comparing results
  found with and without using  the OPP scheme to select the sturmians
  that  form  the  expansion  set.  Note that  the  actual  matrix  of
  potentials is the same throughout though extra information on single
  nucleon  plus a  core  nucleus state  underlying each  sub-threshold
  bound  and resonance  in the  compound system  has been  obtained by
  taking the zero deformation limit.

  Full details of the MCAS  scheme have been published \cite{Am03} and
  the reader is  referred there for those, as  well as for specifics
  of  the notation  we use  herein. In  momentum  space for
  potential matrices $V_{cc'}(p,q)$, one seeks the solution of coupled LS
  equations [see Eq.~(1) in Ref. \cite{Am03}], which involve both open
  and  closed channel  contributions.  With incident  energy $E$,  the
  channel wave numbers for the  open and closed channels are $k_c$ and
  $h_c$ respectively. Solutions of those LS equations are sought using
  expansions  of the  potential matrix  elements in  (finite)  sums of
  energy-independent                  separable                 terms,
  \begin{equation}
    V_{cc'}(p,q) \sim  \sum^N_{n = 1} \hat{\chi}_{cn}(p)
    \eta^{-1}_n \hat{\chi}_{c'n}(q)\; ,
    \label{finiteS}
  \end{equation}
  where $\hat{\chi}_{c'n}(q)$ are the Fourier-Bessel transforms of the
  selected  sturmians  whose  eigenvalues  are  $\eta_n$.  To  predict
  observables one  requires the  multichannel $S$-matrix. In  terms of
  the  multi-channel  $T$-matrix,   that  has  closed  algebraic  form
  \begin{eqnarray}
    S_{cc'} & = & \delta_{cc'} -i \pi \mu \sqrt{k_c k_{c'}}\; T_{cc'}
    \nonumber \\
    T_{cc'} & = & \sum_{n,n' = 1}^N 
    \hat{\chi}_{cn}(k_c) \left([\mbox{\boldmath $\eta$} - \mathbf{G}_0]^{-1}
    \right)_{nn'}\ \hat{\chi}_{c'n'}(k_{c'})\ ,
    \label{multiS}
  \end{eqnarray}
  where   now  $c,c'$   refer   to  open   channels   only.  In   this
  representation,  $\mathbf{G}_0$  and  \mbox{\boldmath  $\eta$}  have
  matrix elements
  \begin{align}
    \left[ \mathbf{G}_0 \right]_{nn'} & = \mu \left[
    \sum^{\text{open}}_c \int^{\infty}_0 \hat{\chi}_{cn}(x)
    \frac{x^2}{k^2_c - x^2 +i\varepsilon} \hat{\chi}_{cn'} \, dx
    \right. \nonumber \\
    & \phantom{= \mu} - \left. \sum^{\text{closed}}_c \int^{\infty}_0
    \hat{\chi}_{cn}(x) \frac{x^2}{h^2_c + x^2} \hat{\chi}_{cn'}(x) \,
    dx \right] \; , \nonumber \\
    \left[ \mbox{\boldmath $\eta$} \right]_{nn'} & = \eta_n \delta_{nn'}
    \; .
    \label{xiGels}
  \end{align}
  The bound states of the compound  system are defined by the zeros of
  the matrix determinant when the  energy $E$ is negative, and so link
  to     the      zeros     of     $\{      \left|     \mbox{\boldmath
  $\eta$}-\mathbf{G}_0\right|    \}$     when    all    channels    in
  Eq.~(\ref{xiGels}) are closed.

  As  noted   above  the   sturmians  are  solutions   of  homogeneous
  Schr\"odinger equations for the  matrix of potentials. In coordinate
  space if those potentials  are designated by local forms $V_{cc'}(r)
  \delta(r-r')$, the OPP method is to use sturmians that are solutions
  for                        nonlocal                       potentials
  \begin{equation}
    \mathcal{V}_{cc'}(r,r')  =  V_{cc'}(r)\delta(r-r')  +  \lambda
    A_c(r) A_c(r')\delta_{cc'} ,
  \end{equation}
  where $A(r)$ is  the radial part of the  single particle bound state
  wave function  in channel $c$  spanning the phase space  excluded by
  the Pauli principle. The OPP  method holds in the limit $\lambda \to
  \infty$, but use of $\lambda = 100$~MeV suffices.

  The  spectrum of  $^{12}$C also  was calculated  in the  shell model
  using the  program OXBASH \cite{Ox86} and with  the MK3W interaction
  \cite{Wa89}. The positive parity  states of $^{12}$C were calculated
  in a complete $(0+2)\hbar\omega$ space using this interaction, while
  the  negative   parity  states  were  calculated   in  a  restricted
  $(1+3)\hbar\omega$  space.  In  both  calculations the  same  single
  particle  basis of $0s$  up to  and including  the $0f1p$  shell was
  used. Hence the restriction  from a full $(1+3)\hbar\omega$ study is
  that we have not included  the $0g1d2s$ shell. With exceptions, most
  notably the super-deformed $0^+_2$  state at 7.654~MeV and the known
  collective $3^-$  state at 9.64  MeV, the calculated spectrum  to 20
  MeV excitation agrees well  with observation \cite{Am00}. So also do
  results  of  calculations   \cite{Am00}  of  elastic  and  inelastic
  scattering   data  (form  factors   from  electron   scattering  and
  differential  cross  sections   and  analyzing  powers  from  proton
  scattering)  without the  need  for any  \textit{a posteriori}  core
  polarization corrections.

  Of interest here  are the details of the  low lying spectrum. First,
  in Table \ref{C12occs}, we list the nucleon shell occupancies in the
  three lowest states of $^{12}$C.  Clearly the $0s$- and $0p$- shells
  have dense occupancy: essentially  4 nucleons filling the $0s$ shell
  while there  are almost  8 nucleons in  the $0p$ shell.  Those eight
  nucleons  are distributed  between the  sub-shells, so  blocking the
  $0p_{\frac{3}{2}}$  orbit as we  do in  using the  OPP method  is an
  approximation. Note also that  the second excited state of $^{12}$C,
  the  $0^+_2$(7.654  MeV)  is  well  known  to  be  a  super-deformed
  $3\alpha$ chain  and, as such, a  much larger space is  needed for a
  good                                                    description.
  \begin{table}
    \caption{\label{C12occs}   Shell   occupancies   of  protons   (or
      neutrons) in states of $^{12}$C.}
    \begin{ruledtabular}
      \begin{tabular}{cccc}
	orbit & $0^+_1$ & $2^+_1$ & $0^+_2$\\
	\hline
	$0s_{\frac{1}{2}}$ & 1.963 & 1.962 & 1.968\\
	$0p_{\frac{3}{2}}$ & 3.054 & 2.858 & 3.075\\
	$0p_{\frac{1}{2}}$ & 0.842 & 1.028 & 0.804\\
	higher orbits   & 0.124 & 0.129 & 0.120
      \end{tabular}
    \end{ruledtabular}
  \end{table}
  The lowest three $0^+$ states in  our shell model are the ground, at
  12.25, and  at 23.03~MeV.  They are much  more spread  than measured
  energies               and               have              structure
  \begin{align*}
    \left| {}^{12}\text{C} \left( 0^+_1 \right) \right\rangle & =
    80.525\% \left| 0\hbar\omega \right\rangle + 19.475\% \left|
    2\hbar\omega \right\rangle \\ 
    \left| {}^{12}\text{C} \left( 0^+_2 \right) \right\rangle & =
    78.213\% \left| 0\hbar\omega \right\rangle + 21.786\% \left|
    2\hbar\omega \right\rangle \\
    \left| {}^{12}\text{C} \left( 0^+_3 \right) \right\rangle & =
    \; \: 9.066\% \left| 0\hbar\omega \right\rangle + 90.934\% \left|
    2\hbar\omega \right\rangle \; .
  \end{align*}
  Notice  that the  first  dominantly $2\hbar  \omega$  state lies  at
  23.03~MeV excitation;  a calculated energy  that can be  expected to
  fall with the addition of higher $\hbar \omega$ components. That has
  not been  seen sufficiently  at least to  the $4\hbar  \omega$ level
  with  an \textit{ab initio}  shell model  \cite{Na00}. So  while the
  $0^+_3$ state may  be the one that is  super-deformed, a convergence
  in energy will require a greatly increased space. But as the $0^+_2$
  state is not very important in the formation of resonances and bound
  states \cite{Am03},  use of our  shell model should suffice  for the
  comparisons we make. This may have more bearing on future results as
  we  move  to  use  a  microscopic  MCAS in  which  the  matrices  of
  interaction potentials will be formed using nucleon-based structure.

  Next we consider how each state in  the low excitation spectrum
  of $^{13}$C  maps onto a  single neutron added  to any of  the three
  selected  states of  $^{12}$C. The  relevant  one-body spectroscopic
  amplitudes         for         $I^\pi         \to         J^{\pi'}$,
  \begin{equation}
    S_{j  \frac{1}{2}} = \frac{1}{\sqrt{2(2J+1)}}  \left\langle \left(
    {}^{13}\text{C}      \right)      J^{\pi'}      \left|\left|\left|
    a^{\dag}_{j,\frac{1}{2}}        \right|\right|\right|       \left(
    {}^{12}\text{C} \right) I^{\pi}_i \right\rangle \; ,
  \end{equation}
  are listed in Table~\ref{C13-C12}.
  \begin{table}
    \caption[]{\label{C13-C12}  Dominant  components  of  shell  model
      spectroscopic amplitudes. Energies in brackets are in MeV.}
    \begin{ruledtabular}
      \begin{tabular}{cclrlrlr}
	\multicolumn{2}{c}{$^{13}$C} & \multicolumn{6}{c}{$^{12}$C} \\
	& & \multicolumn{2}{c}{$0^+_1$} & \multicolumn{2}{c}{$2^+_1$}
	& \multicolumn{2}{c}{$0^+_2$} \\
	\hline
	$\frac{1}{2}^-$ & (g.s.) & $0p_{\frac{1}{2}}$ & $-$0.7285 & 
	$0p_{\frac{3}{2}}$ & $-$1.0040 & $0p_{\frac{1}{2}}$ & $-$0.4738
	 \\
	$\frac{1}{2}^+$ & (3.09) & $1s_{\frac{1}{2}}$ & $-$0.9088 & 
	$0d_{\frac{5}{2}}$ & $-$0.3162 & $1s_{\frac{1}{2}}$ & $-$0.0605
	 \\ 
	$\frac{3}{2}^-$ & (3.68) & $0p_{\frac{3}{2}}$
	 & 0.4504 &  $0p_{\frac{3}{2}}$ & $-$1.0040 &
	 $0p_{\frac{3}{2}}$ & $-$0.3284 \\ 
	 & & & & $0f_{\frac{5}{2}}$ & $-$0.8342 & &  \\
	$\frac{5}{2}^+$ & (3.85) & $0d_{\frac{5}{2}}$ & 0.8129 &
	$0d_{\frac{5}{2}}$ & 0.4799
	 & $0d_{\frac{5}{2}}$ & 0.0096  \\
	 & &  & & $0d_{\frac{3}{2}}$ & $-$0.1361 & & \\
	  & & & & $1s_{\frac{1}{2}}$ & 0.0840 & & \\
	$\frac{5}{2}^+$ & (6.86) & $0d_{\frac{5}{2}}$ & $-$0.2147 &
	 $0d_{\frac{5}{2}}$ & 0.5372 & $0d_{\frac{5}{2}}$ & $-$0.0102 \\ 
	  & & & & $0d_{\frac{3}{2}}$ & $-$0.0907 & &  \\
	  & & & & $1s_{\frac{1}{2}}$ & $-$0.7714 & & \\
	$\frac{5}{2}^+$ & (8.88) & $0d_{\frac{5}{2}}$ & $-$0.0349 & 
	$0d_{\frac{5}{2}}$ & $-$0.2568 & $0d_{\frac{5}{2}}$
         & 0.2829 \\
	  & & & & $0d_{\frac{3}{2}}$ & $-$0.2694 & & \\
	 & &  & & $1s_{\frac{1}{2}}$ &  $-$0.2391 & & 
      \end{tabular}
    \end{ruledtabular}
  \end{table}
  The shell model  calculations gave more values for  addition of that
  neutron in  higher shell states, but  those spectroscopic amplitudes
  (not listed) all have magnitudes less than 0.1.

  Results  of  calculations of  the  neutron-$^{12}$C system  reported
  previously \cite{Am03}, used the parameter values that are specified
  in                       Table                       \ref{OMparams}.
  \begin{table}
    \begin{ruledtabular}
      \caption{\label{OMparams} Parameter values of the base potential
      (in MeV).}
	  \begin{tabular}{cccccc}
	    & $V_0(\pi)$ & $V_{\ell \ell}(\pi)$ & $V_{\ell s}(\pi)$ &
	    $V_{Is}(\pi)$ \\ 
	    \hline
	    $\pi = -$ & $-$49.144 & 4.559 & 7.384 & $-$4.770 \\
	    $\pi = +$ & $-$47.563 & 0.610 & 9.176 & $-$0.052 \\
	    \hline
	    Geometry &  $R_0 = 3.09$ fm  & $a = 0.65$ fm  & $\beta_2 =
	    -0.52$ & 
	  \end{tabular}
    \end{ruledtabular}
  \end{table}

\begin{figure}
    \scalebox{0.4}{\includegraphics*{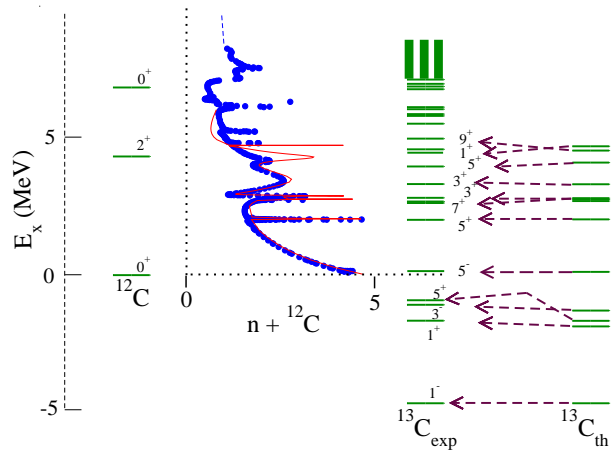}}
    \caption[]{\label{n-12C-fig}  The spectra  of $^{12,13}$C  and the
      elastic  cross section  (barn) for  n+$^{12}$C system.  The ENDF
      data  (circles)  \cite{ENDF} are  compared  with  our full  MCAS
      results  (solid  line).  Note  that the  identification  of  the
      $^{13}$C states' spin-parities is $2J^\pi$.}
  \end{figure}

  In Fig.  \ref{n-12C-fig}, the results  are compared with  data, both
  elastic  scattering  cross  sections  as  well as  the  spectrum  of
  $^{13}$C.  Therein it  is clear  that the  three states  of $^{12}$C
  suffice to deal with information  to $\sim 10$~MeV excitation in the
  compound  with  corroboration in  the  scattering  of  up to  5~MeV.
  Spin-parity   assignments,  bound   state  energies   and  resonance
  centroids, widths  of the resonances, and  the background scattering
  all    are    very     well    matched    by    the    calculations.
  \begin{table}
    \begin{ruledtabular}
      \caption{\label{Zero-b+Vss}  Pauli effects on  sub-threshold and
	bound states  in the  continuum in the  limit $\beta_2  \to 0$
	(with $V_{ss} = 0$).}
      \begin{tabular}{ccccc}
	$J^\pi$ & With Pauli & No Pauli & n+${}^{12}$C \\
	\hline
	$\frac{1}{2}^+$ & - &  $-$26.57 & $0s_{1/2} + 0^+_1$ \\
	$\frac{3}{2}^+, \frac{5}{2}^+$ & - &  $-$22.13 & $0s_{1/2} +
	2^+_1$ \\ 
	$\frac{1}{2}^+$ & - &  $-$18.91 & $0s_{1/2} + 0^+_2$ \\
	$\frac{3}{2}^-$ & - &  $-$8.849 & $0p_{3/2} + 0^+_1$ \\
	$\frac{1}{2}^-$ & $-$4.685 & $-$4.685 & $0p_{1/2} + 0^+_1$ \\ 
	$\frac{1}{2}^-, \frac{3}{2}^-, \frac{5}{2}^-, \frac{7}{2}^-$ &
	- & $-$4.410 &  $0p_{3/2} + 2^+_1$ \\
	$\frac{3}{2}^-$ & - & $-$1.195 & $0p_{1/2} + 0^+_2$ \\
	$\frac{1}{2}^+$ & $-$0.837 & $-$0.837 & $1s_{1/2} + 0^+_1$ \\
	$\frac{3}{2}^-, \frac{5}{2}^-$ & $-$0.246 & $-$0.246 &
	$0p_{1/2} + 2^+_1$ \\ 
	$\frac{5}{2}^+$ & $-$0.171 &  $-$0.171 & $0d_{5/2} + 0^+_1$ \\
	$\frac{1}{2}^-$ &  \phantom{$-$}2.969 & \phantom{$-$}2.969 &
	$0p_{1/2} + 0^+_2$ \\ 
	$\frac{3}{2}^+, \frac{5}{2}^+$ &  \phantom{$-$}3.601 &
	\phantom{$-$}3.601 &  $1s_{1/2} + 2^+_1$ \\
	$\frac{1}{2}^+, \frac{3}{2}^+, \frac{5}{2}^+, \frac{7}{2}^+,
	\frac{9}{2}^+$ & \phantom{$-$}4.267 & \phantom{$-$}4.267 &
	$0d_{5/2} + 2^+_1$
      \end{tabular}
    \end{ruledtabular}
  \end{table}
  A most interesting  feature is what occurs as  the coupling tends to
  zero. In  that limit,  all of the  compound resonances shrink  to be
  bound states in the continuum. In this limit, calculations were made
  with  the  spin-spin  interaction  strengths  set to  zero,  and  so
  offsetting a splitting that is  most evident with the two odd parity
  states built  from coupling  with a $0p_{\frac{1}{2}}$  neutron. The
  results  of   these  limit  calculations  are   collected  in  Table
  \ref{Zero-b+Vss}. Therein  the states are  listed in the  order from
  most  bound to  largest continuum  energy whether  they are  real or
  spurious. The  energies listed in columns~2 and  3 respectively were
  found  in  the zero  deformation  limit  with  and without  the  OPP
  treatment  of Pauli  blocking. In  the last  column we  display what
  dominant character  (neutron orbit coupled to state  in $^{12}$C) is
  found for  each state in $^{13}$C. Disregarding  the Pauli principle
  clearly gives  many spurious states. However, notice  that there are
  matching  entries  for  every  resonance  state  whether  the  Pauli
  principle  is  taken into  account  or not.  That  has  led to  the
  erroneous belief that a simple adjustment of parameter values is all
  that  is needed  to define  scattering cross  sections and  that the
  Pauli principle effects are  unimportant for scattering. Not only is
  that phenomenology  not guaranteed to  work in other cases  but also
  the  mixing of  components  caused by  finite  deformation is  quite
  different  when  the Pauli  principle  is  or  is not  satisfied.  A
  calculation  made ignoring  the Pauli  principle gives  an incorrect
  description of all states.

  The  resonance centroids  tend to  three limits.  The highest  is at
  4.267~MeV with five  entries from $\frac{1}{2}^+$ to $\frac{9}{2}^+$
  as is formed by attaching a $0d_{5/2}$ neutron to the $2^+$ state in
  $^{12}$C. The second is at 3.601~MeV having two entries which equate
  to a  $1s_{1/2}$ neutron coupled to  the $2^+$ state  of the target.
  The  third, the  only odd  parity resonance  ascertained  from these
  calculations within the range of energies to 5~MeV, is identified as
  a $\frac{1}{2}^-$  resonance. It lies 7.65~MeV  above the calculated
  value for the $^{13}$C ground  state and can then be associated with
  binding a $0p_{1/2}$ neutron to the second $0^+$ state of $^{12}$C.

  The bound  states are  less clear with  regard to  dominant particle
  coupling   character.    From   shell   model    calculations,   the
  $\frac{1}{2}^-$  (ground state)  and the  $\frac{3}{2}^-$  state are
  sizable mixtures  of $p$-shell nucleon  coupling to both  the ground
  and  $2^+$   states  in   $^{12}$C.  But  the   $\frac{1}{2}^+$  and
  $\frac{5}{2}^+$  bound  states  in  $^{13}$C are  dominantly  formed
  respectively by a $1s_{1/2}$ and a $0d_{5/2}$ neutron coupled to the
  ground state  of $^{12}$C. The  energies found in the  zero coupling
  limit  support  the  inferences   made  above.  Notably,  the  bound
  $\frac{5}{2}^+$  tends to $-0.171$~MeV  in that  limit, as  $4.267 -
  (-0.171) = 4.438$~MeV;  the excitation energy of the  $2^+$ state in
  ${}^{12}$C.  Likewise  the  doublet  $\frac{3}{2}^+,  \frac{5}{2}^+$
  tends to $3.601$~MeV and as the bound state $\frac{1}{2}^+$ tends to
  $-0.837$~MeV  in  the  zero  coupling  limit, $3.601  -  (-0.837)  =
  4.438$~MeV,  the excitation energy  of the  $2^+$ state  in $^{12}$C
  once  more. Finally the  $\frac{3}{2}^-$ and  $\frac{5}{2}^-$ states
  both are  bound by $-0.246$~MeV and  so the energy gap  of that pair
  from the $\frac{1}{2}^-$ state ($-4.685$~MeV), is 4.439~MeV.

  But  deformation makes  significant admixing  of these  nucleon plus
  core nucleus elements. Indeed it is this spurious mixing that is the
  most serious concern  about calculations that do not  take the Pauli
  principle into account.  To reveal that, we have  repeated the limit
  calculations  excising  the  OPP  effects and  thereby  solving  the
  problem in  a way equivalent  to coordinate-space solutions  of such
  coupled  channels  equations.  That  calculation clearly  has  given
  spurious states.

  Summarizing,   the  MCAS   approach  has   been  used   to  evaluate
  (low-energy)  n-$^{12}$C  elastic  scattering  and  to  characterize
  sub-threshold  states of $^{13}$C.  A collective  model prescription
  with the three lowest states  in the $^{12}$C spectrum was used. The
  results well  match observed  data but only  when allowance  for the
  influence of  the Pauli principle was made.  Without such allowance,
  many spurious  states result. Most  strikingly, the ground  state of
  $^{13}$C then has the wrong  spin-parity and a binding far in excess
  of the known  value. But more disturbing is that  when states may be
  matched (in  energy and  spin-parity) their underlying  nucleon plus
  $^{12}$C compositions are wrong. By tracking results to the $\beta_2
  \to  0$ limit,  the  dominant parentage  of  each sub-threshold  and
  resonant state in this system has been identified.

  \bibliography{MCAS-PRL}

\end{document}